\documentclass[twocolumn,preprintnumbers,showpacs]{revtex4-1}
\usepackage{amsmath}
\usepackage{amssymb}
\usepackage{graphicx}
\usepackage{color}
\usepackage{float}
\usepackage{sistyle}
\usepackage{subfigure}  
\usepackage{verbatim}   
\usepackage{graphicx}
\usepackage{dcolumn}
\usepackage{bm}
\usepackage[citecolor=blue,filecolor=blue,colorlinks=true,urlcolor=blue]{hyperref}
\usepackage{ulem}

\newcommand{\BAN}{\ensuremath{B_{1g}\,}}

\newcommand{\Ts}{\ensuremath{T^{\ast}\,}}
\newcommand{\Tc}{\ensuremath{T_{\rm c}\,}}

\graphicspath{{images/}{../images/}}

\begin{document}

\title{Unambiguous connection between the Fermi surface topology and the pseudogap in Bi$_{2}$Sr$_{2}$CaCu$_{2}$O$_{8+\delta}$ }

\author{B. Loret$^{1}$, Y. Gallais$^1$, M. Cazayous$^1$, R. D. Zhong$^2$, J. Schneeloch$^2$, G. D. Gu$^2$, A. Fedorov$^4$, T. K. Kim$^5$, S. V.Borisenko$^4$ and A. Sacuto$^1$, }

\affiliation{$^1$ Laboratoire Mat\'eriaux et Ph\'enom$\grave{e}$nes Quantiques (UMR 7162 CNRS), Universit\'e Paris Diderot-Paris 7, Bat. Condorcet, 75205 Paris Cedex 13, France\\
$^2$ Matter Physics and Materials Science, Brookhaven National Laboratory (BNL), Upton, NY 11973, USA,\\
$^3$ Laboratoire de Physique des Solides, CNRS, Univ. Paris-Sud, Universit\'e Paris-Saclay, 91405 Orsay Cedex, France,\\
$^4$ IFW-Dresden, Helmholtzstrasse 20, 01069 Dresden, Germany,\\
$^5$ Diamond Light Source, Harwell Campus, Didcot, OX11 0DE, UK.}
\date{\today}


\begin{abstract}

We study the behavior of the pseudogap in overdoped Bi$_{2}$Sr$_{2}$CaCu$_{2}$O$_{8+\delta}$ by electronic Raman scattering (ERS) and angle-resolved photoemission spectroscopy (ARPES) on the same single crystals. Using both techniques we find that, unlike the superconducting gap, the pseudogap related to the anti-bonding band vanishes above the critical doping  $p_c=0.22$. Concomitantly, we show from ARPES measurements that the Fermi surface of the anti-bonding band is hole-like below $p_c$ and becomes electron-like above $p_c$. This reveals that the appearance of the pseudogap depends on the Fermi surface topology in Bi$_{2}$Sr$_{2}$CaCu$_{2}$O$_{8+\delta}$,  and more generally, puts strong constraint on theories of the pseudogap phase.

\end{abstract}


 \maketitle

Revealing the true nature of the pseudogap (PG) phase remains one of the main challenges for understanding physics of hole doped copper oxide superconductors. After several decades of research, key elements emerge without being still able to identify the origin of the PG. The PG sets below a characteristic temperature \Ts $(p)$ which is a decreasing function of the doping $p$. It manifests by a loss of low-energy spectral weight in many different probes \cite{Alloul1989,Warren1989,Homes1995,Opel2000,Loram2001,Kordyuk2002,Bernhard2008,Kohsaka2008,Chatterjee2011,Vishik2012, Sacuto2013,Sakai2013,Benhabib15} with electronic broken-symmetry states \cite{Fauque2006,Daou2009,Shekhter2013,Fujita2014,Badoux2016,Sato2017}. There is now an increasing number of compelling experimental evidences that the PG develops also below the superconducting transition temperature \Tc \cite{Loram2001,Zheng2005,Kondo2009,Shekhter2013,Fujita2014,ManginThro2014,He2014,Hashimoto2015,Loret2016,Badoux2016,Loret2017}.


Rather than trying to determine the origin of the PG phase which remains a difficult task, we propose an alternative approach which consists in finding the conditions of its survival. In our previous investigations, we showed the PG phase collapses abruptly in the overdoped regime of Bi$_{2}$Sr$_{2}$CaCu$_{2}$O$_{8+\delta}$ (Bi-2212) close to $p_c=0.22$ and its end draws a vertical line under the superconducting dome in the $T-p$ phase diagram \cite{Benhabib15,Loret2017}. This observation has to be related to the anisotropy disappearance of the scattering rate reported earlier at the same doping\cite{venturini2002}. We showed the PG collapse coincides with a sharp peak in the density of states (DOS) of the underlying band structure around the anti-nodal region of the Brillouin zone \cite{Benhabib15}. By using ARPES data obtained from Bi-2212 at high doping levels which observed a change of the Fermi surface (FS) topology \cite {Kaminski2006}, we have interpreted the enhancement of the DOS as the manifestation of a doping induced Lifshitz quantum phase transition wherein, as a van Hove singularity crosses the chemical potential, the active hole-like anti-bonding Fermi surface of bilayer Bi-2212 becomes electron-like while the bonding band remains hole-like \cite{Benhabib15}. 
These results suggested indirectly that the PG end in Bi-2212 is due to a change of the FS topology. Note that the Raman response is predominantly sensitive to the anti-bonding band since the latter is close to a density of states singularity which strongly enhances the Raman response as reported in our previous investigations \cite {Benhabib15}. It is generally believed then that the information about the possible existence of a pseudogap in the bonding band is preempted by the large anti-bonding band response.

A possible link between the PG and the Fermi surface topology in cuprates was also inferred from previous works on La$_{2-x}$Sr$_x$CuO$_4$ (LSCO) \cite{Ino2002,Yoshida2006,Park2013} and its related compounds Nd$_{0.4}$Sr$_{x}$CuO$_{4}$ (Nd-LSCO) \cite{Matt2015} and  Bi$_2$Sr$_2$CuO$_{6+\delta}$ (Bi-2201) \cite{Takeuchi2005,Kondo2005}.
 
More recently, a detailed study on low-temperature high-magnetic-field transport measurements under pressure in Nd-LSCO system infers that the PG cannot open on an electron-like Fermi surface \cite{Doiron2017}.  This is confirmed by two recent theoretical studies in the framework of a two-dimensional Hubbard model. They showed that a PG only opens on hole-like Fermi surfaces for a wide range of band structure parameters, even in the strong-coupling regime where the anti-ferromagnetic correlations responsible for the PG are short ranged \cite{Wu2017,Braganca2017}.

In order to demonstrate unambiguously that the change of the FS topology actually exists, it is not an accident and plays indeed, a key role in the disappearance of the PG in Bi-2212, we have combined ARPES and ERS measurements on the same single crystals at high doping levels to check that both the ERS and ARPES signatures of the PG disappear simultaneously when the anti-bonding band of the FS becomes electron-like. 

Our studies are focused on two Bi-2212 overdoped single crystals OD 47K (p=0.236) and OD 77K (p=0.201) respectively located below and above the Lifshitz transition which is located close to $p_c=0.22$. \\
The doping level was estimated from the pair breaking-peak location detected in the Raman spectra \cite{Note}. The doping estimate from ARPES (although less accurate due to many diffraction replicas) is consistent with that of Raman. We find from the Luttinger's theorem: $p=0.24 \pm 0.01$ for the OD 47K compound and $p=0.21\pm 0.01$ for the OD 77K compound. 


Raman experiments have been carried out using a triple grating spectrometer (JY-T64000) equipped with a liquid-nitrogen-cooled CCD detector. The 532 nm laser excitation line was used from a diode pump solid state laser. The \BAN (anti-nodal) geometry has been obtained from cross polarizations at 45$^o$ from the Cu-O bond directions. We  got an accuracy on the crystallographic axes orientation with respect to the polarizors close  to $2^o$. All the spectra have been corrected for the Bose factor and the instrumental spectral response. 
Measurements below and above \Tc have been performed using an ARS closed-cycle He cryostat. The laser power at the entrance of cryostat was maintained below 2 mW to avoid over heating of the crystal estimated to 3 K/mW at 10 K. The crystals were cleaved before measurements.\\

ARPES measurements were performed at Diamond Light Source at the I05 beamline \cite{Hoesh2017} as well as at BESSY II (Helmholtz Zentrum Berlin) at the UE112-PGM2b beamline using the "1$^3$-ARPES" end-station \cite{Borisenko2012}. In both cases crystals were cleaved in ultra-high vacuum to expose mirror-like surfaces, energy resolution was kept below 5 meV and photon energies between 20 eV and 100 eV were used. The photon energy has been carefully selected each time when the intensity of the particular band should have been enhanced in comparison with other features. The photon energy range (50-55 eV) has been selected in order to probe mostly the anti-bonding band.  For the highest resolution measurements the samples were cooled down to the lowest achievable temperatures, i.e. 8 K at Diamond and 1 K at BESSY. The crystals were first measured by Raman and then by ARPES.

Our first venture was to check the consistency of the ARPES and ERS measurements above $p_c$. 

In the first row of Figure 1, are displayed the anti-nodal ERS response and the energy distribution curves (EDCs) related to the anti-bounding band at the anti-nodes (AN) and the nodes (N) of the OD 47K compound (p=0.236) above and below \Tc. 
Because ERS spectroscopy is a two particles probe (both occupied and unoccupied states are involved in the ERS process), the energy of the pair breaking peak in the superconducting (SC) Raman spectrum (red/grey curve in Fig.1 (a)) is at twice the SC gap energy. We find $\Delta_{SC} = 9.2 \pm 0.2$ meV. On the other hand, ARPES is a one-particle probe: solely occupied states are detected. The estimate of the SC gap by ARPES can be deduced from the EDCs in two different ways. Either by measuring the difference in energy (i) between the quasi-particles peaks below and above \Tc or (ii) between the quasi-particle peaks at the nodes (N) and at the anti-nodes (AN) below \Tc. The EDCs are sufficiently sharp to extract the gap from the peak location directly, without using additional data processing, such as  symmetrization.
In case (i) (Fig. 1 (b)) we find $\Delta_{SC1}=9\pm 1$ meV very close to the SC gap from ERS. In case (ii) (Fig.1 (c)), the SC gap value  $\Delta_{SC2}$ is slightly lower, $8 \pm 1$ meV. In the light of our uncertainties on these two methods, it is difficult  to say whether this difference in the determination of the SC gap is significant or not and further investigations are required to answer this question. 

 

\begin{figure}[tb!]
\begin{center}
\includegraphics[width=9cm,height=8cm]{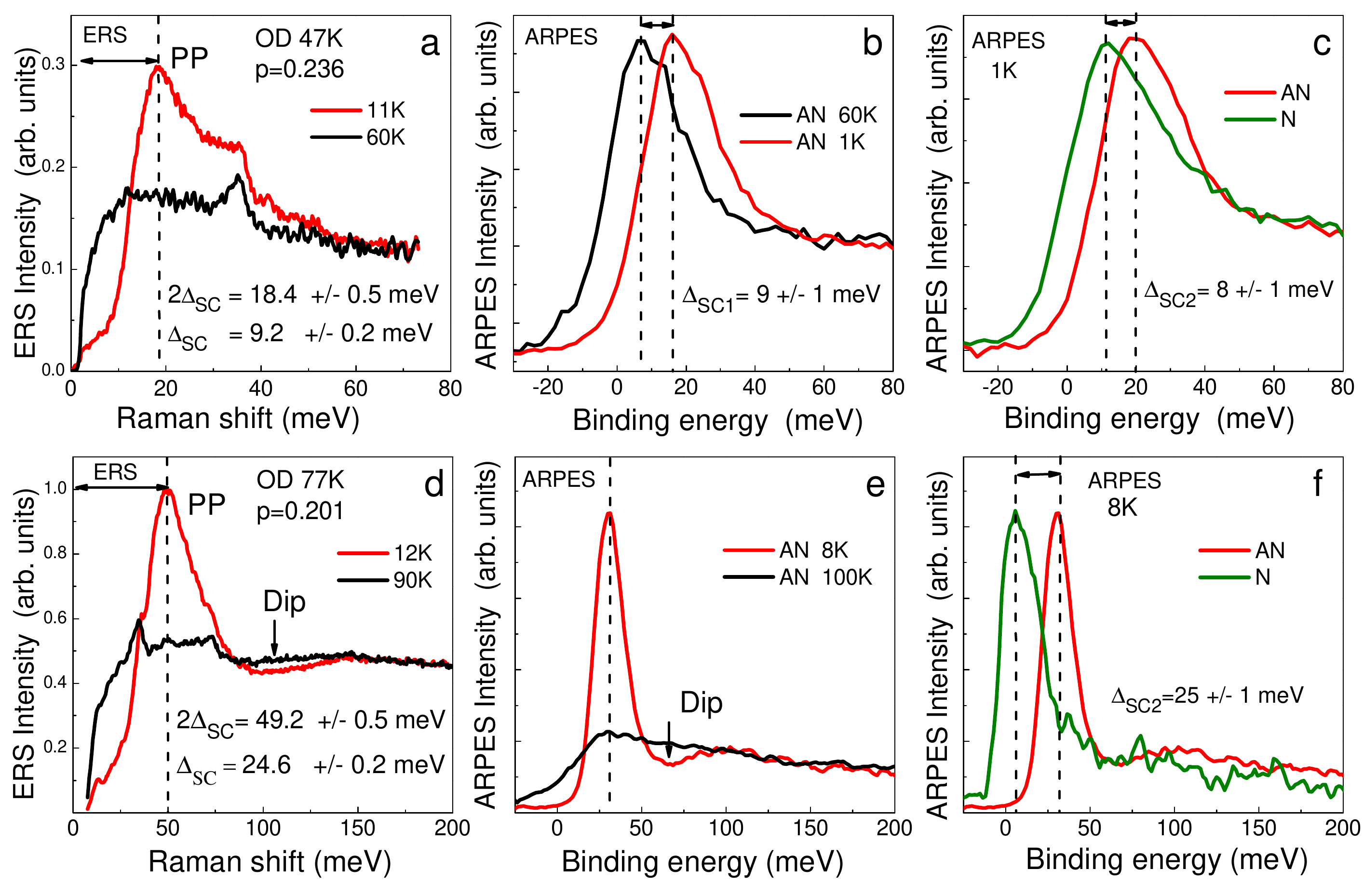}
\caption{(Color online). First row: ERS and ARPES spectra of an Bi-2212 overdoped (OD 47K) crystal above $p_c=0.22$, (a) \BAN (anti-nodal) Raman response, (b) Anti-nodal EDCs of the anti-bonding band below and above \Tc  and  (c) Anti-nodal and Nodal EDCs of the anti-bonding band below \Tc. Note that the EDCs have been selected to emphasize momentum and temperature dependence respectively. The anti-nodal EDC from Fig.1(c) is a little broader than the one in Fig.1 (b) and this caused a slightly higher binding energy of the peak. For the gap determination this difference plays no role since it is relative changes of the line-shape which are important. Second row: ERS and ARPES spectra of Bi-2212 overdoped (OD 77K) crystal below $p_c$, (d) Anti-nodal Raman response, (e) Anti-nodal EDCs of the anti-bonding band  below and above \Tc and (f) Anti-nodal and Nodal EDCs of the anti-bonding band below \Tc. The peak at 35 and weak reminiscent peaks at 50 and 75 meV observed both in (a) and (d) are phonon lines\cite{Liu1992,Hewitt1999}. Zero binding energy corresponds to the Fermi level.}
\label{fig:1}
\end{center}\vspace{-7mm}
\end{figure}

In the second row are reported the Raman response function and the ARPES EDCs of the anti-bonding band related to the  OD 77K compound (p=0.201).
The SC gap $\Delta_{SC} =25$ meV deduced from the pair breaking peak location (Fig. 1(d)) and the energy difference between the N and AN quasi-particle peaks $\Delta_{SC2}$ (Fig.1 (f)) are in remarkably good agreement. Interestingly, there is no way to define the SC gap from the ARPES EDCs above and below \Tc because the PG appearing in the normal state (see Fig.1 (e)) pushes the quasi-particles peak to the same binding energy as the Bogoliubov quasi-particle peak. This clearly contrasts with the EDCs of the OD 47K compound (Fig.1 (b)) where the quasi-particles peaks below and above \Tc are located at different energies.  This is the first evidence from our ARPES data that the PG exists at p=0.201 below $p_c\approx 0.22$. The second proof comes from the ERS data. In Fig.1 (d), on the right energy side of the pair breaking peak (PP), a dip in the SC electronic continuum is observable (see Fig.1 (d)), whereas no dip is detected in the SC Raman spectrum of Fig.1 (a). In previous works, we showed this PP-dip structure results from the interplay between the PG and the SC gap, and can be smoothly connected to the PG appearing in the electronic spectrum above \Tc \cite{Loret2016,Loret2017}. Note that a dip in the EDC of OD 77K in Fig.1 (e) is also detected while no dip is observed in the EDC of OD 47K in Fig.1 (b). The origin of the peak-dip and hump detected by ARPES is still under debate. Several scenarios have been proposed: strong electron-boson coupling, bilayer splitting effect, PG effect \cite{Eschrig2000,Kordyuk2002a,Borisenko2003, Damascelli2003,Cuk2004,Hashimoto2015,Mou2017}. In view of our results the PG could also be considered as a possible origin of the peak-dip-hump structure detected in ARPES.

\begin{figure}[tb!]
\begin{center}
\includegraphics[width=7cm,height=7cm]{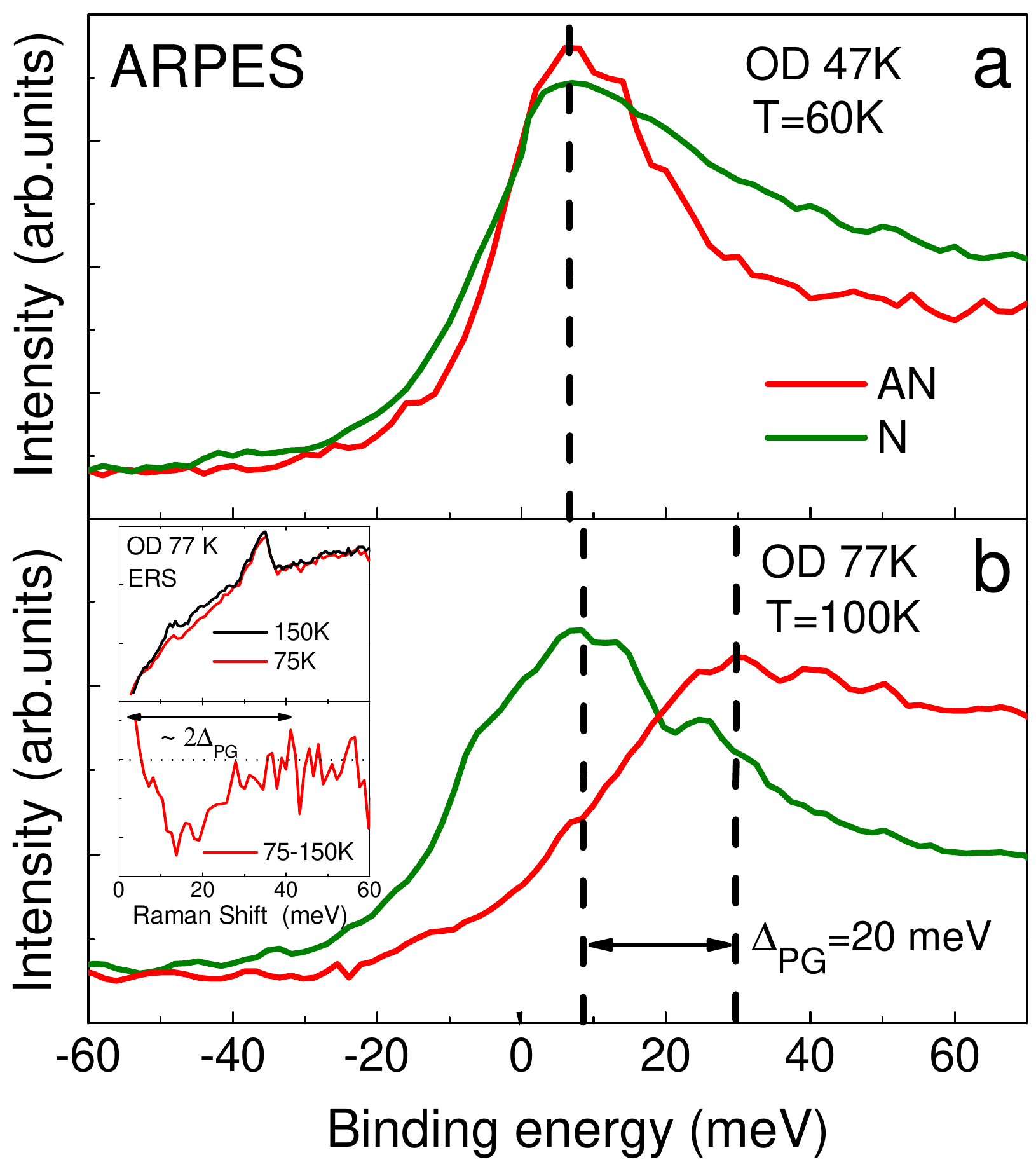}
\caption{(Color online). Anti-bonding band energy distribution curves at the Nodes and Antinodes of the overdoped Bi-2212. (a) OD 47K and (b) OD 77K compound in the normal state.  The inset in (b) displayed (up) the electronic Raman background measured just above \Tc (75K) and at \Ts (150K) for the OD 77K compound; (bottom) the electronic background at 75 K subtracted from the one at 150 K. The energy size of the depletion is approximately 40 meV corresponding (after division par 2) to $\Delta_{PG}\approx20$ meV. The peak around 35 meV is a phonon line. }
\label{fig:2}
\end{center}\vspace{-7mm}
\end{figure}

Finally, we bring a third experimental evidence of the presence of the pseudogap below $p_c$ and its absence above by comparing the N and AN EDCs of OD 47K and OD 77K above \Tc. In Fig.2 (a) are displayed the AN (red curve) and N (green curve) measured at T= 60 K. Although the AN and N quasi-particle peaks have different widths, they overlap well. In sharp contrast, the AN and N quasi-particle peak locations related to the OD 77K compound (measured at 100 K) are quite distinct (see Fig. 2 (b)). Indeed close to the Fermi level the PG opens with a loss of low-energy spectral weight at the AN which pushes back the AN peak to higher binding energy in comparison to the N one. The difference in energy (peak to peak) gives $\Delta_{PG}=20\pm 1$ meV which is in the same energy range that the PG gap energy inferred from ERS see Inset of Fig.2 (b).  In the top panel of the inset have been reported the Raman electronic backgrounds of OD 77K at \Tc and \Ts. The subtracted background (bottom of the inset) gives an estimate of the loss of low-energy spectral weight related to the normal state PG. We can then define the energy from which the depletion starts and found to $2\Delta_{PG}=40\pm 5$ meV. On the contrary, no depletion as the temperature decreases is observed in the Raman spectra of overdoped Bi-2212 compounds with a \Tc below 60 K \cite{Benhabib15} and a fortiori for the OD47K compound. 
  
At this step,  our ERS and ARPES investigations show without ambiguity that the PG exists in the Bi-2212 compound at p=0.201 (related to OD 77K) and does not exist anymore at p=0.236 (which corresponds to to OD 47K). These two doping levels are respectively located below and above $p_c\approx 0.22$, the doping level for which we have suggested a change of the Fermi surface topology \cite{Benhabib15,Loret2017}. Now, we prove that change of the FS topology actually happens.
  
In Figure 3, we compare the spectral weight maps in momentum space at the Fermi level of Bi-2212 for the two doping levels p=0.236 (OD 77K) and p=0.201 (OD 47K) carried out at essentially the same experimental conditions. We use photon energy at 26 eV to enhance the emission from the anti-bonding band. Both Fermi surface maps look typical for Bi-2212 showing the diffraction replicas. The latter are the result of the diffraction of the photo-electrons on the top-most layer. It is known that Bi-2212 are approximately 5:1 structurally modulated along the Cu-Cu bonds \cite{Borisenko2000}. Therefore the primary Fermi surface contours will be replicated, i.e. shifted by approximately one fifth of the Brillouin zone along the diagonal in both directions resulting in the similar but weaker features on the map. \\

\begin{figure}[tb!]
\begin{center}
\includegraphics[width=8.5cm,height=9cm]{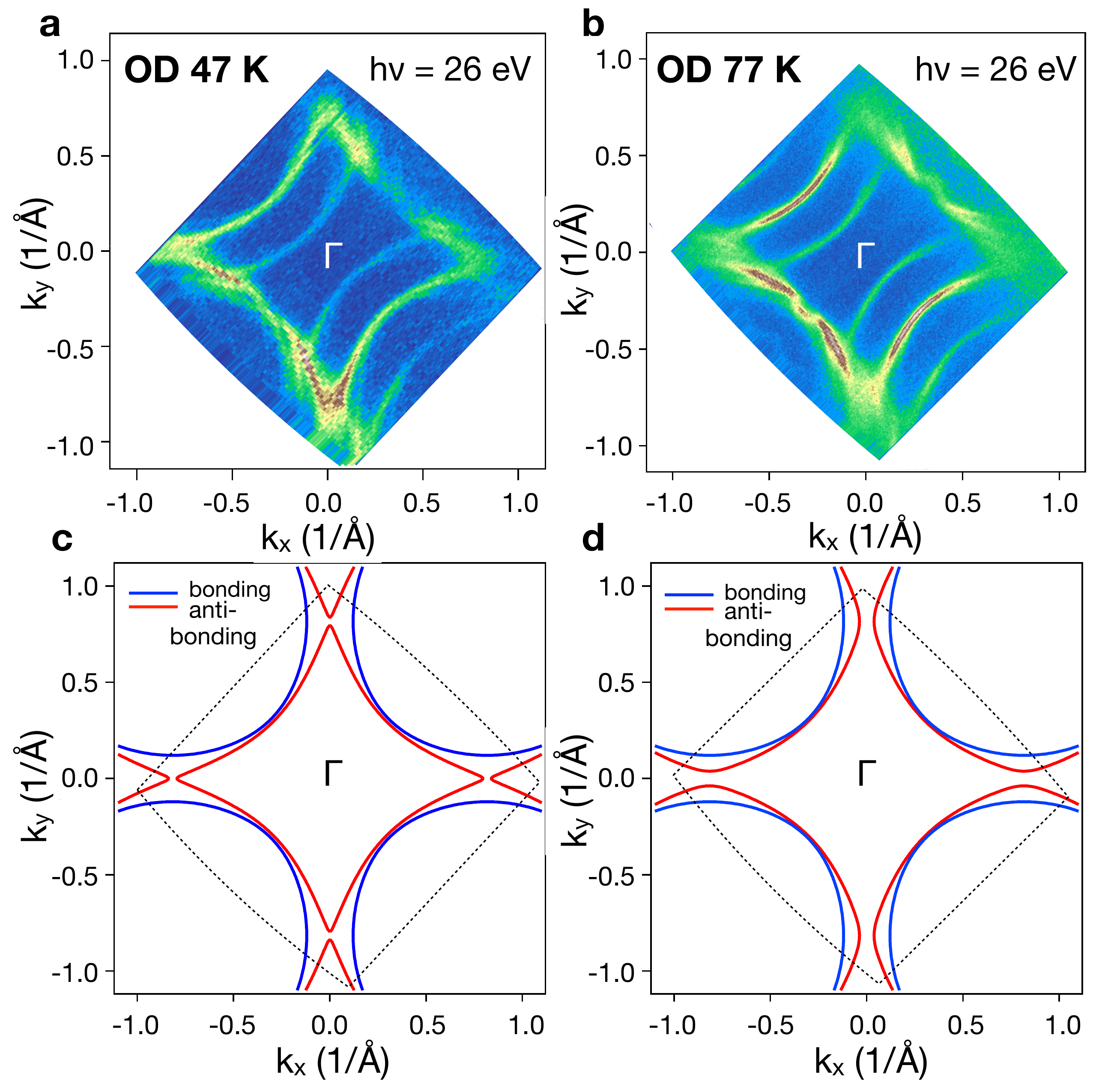}
\caption{(Color online). Fermi surface maps of (a) OD 47K and (b) OD 77K compound taken at 8 K using 26 eV photons. Intensity is integrated within 20 meV window centered at the Fermi level to minimize the influence of the superconducting gap. Results of the tight-binding fit to the experimental data from (c) OD 47K and (d) OD 77K samples. Dotted contour corresponds to the area map in the experiment in panels (a) and (b) correspondingly. Diffraction replicas are not shown. $t_{0,0}$=0.416, $t_{0,1}$=$t_{1,0}$=-0.653 and $t_{1,1}$=0.542 for bonding bands. $t_{0,0}$=0.451 for anti-bonding bands. $t_{1,1}$=0.450, $t_{0,1}$=$t_{1,0}$=-0.630 and $t_{1,1}$=0.463, $t_{0,1}$=$t_{1,0}$=-0.653 for OD 47K and OD 77K anti-bonding bands respectively.}.
\label{fig:2}
\end{center}\vspace{-7mm}
\end{figure}

Already a visual comparison of Fig.3 (a) and Fig.3 (b) shows that the underlying Fermi surfaces are slightly different. The distance between the $\Gamma$ point and the Fermi surface contour along the diagonal is larger in the less hole-doped compound, as expected. The main difference, however, is the contour of the anti-bonding Fermi surface itself. Due to the favorable experimental conditions (photon energy and geometry), this contour is the most intense feature on the maps. In panel (a) this contour is closed around the $\Gamma$ point, whereas it is open in panel (b).\\

Because of the presence of the superconducting gap in the anti-nodal region, both Fermi surface contours become smeared-out on the maps, complicating the comparison. This occurs due to the finite integration energy window (20 meV) of the ARPES signal near the Fermi level, which increases the momentum width of the features following the BCS-like bending-back behavior in the superconducting state.

On the other hand, the spectral function at these low temperatures has the sharpest peaks at all binding energies. Using this advantage, we fitted the simple tight-binding model ($E(k_x, k_y)=t_{0,0} + t_{0,1}cos(y) + t_{1,0}cos(x) + t_{1,1}cos(x)cos(y)$) to both data sets. \\

We used not only the Fermi surface maps shown in Fig.3 but also the underlying dispersion in the broad energy interval below the Fermi level not shown here. The fit procedure (involving the tight-binding function with $k_x,k_y$ and energy as variables) allows us to avoid the immediate vicinity of the Fermi level where the gap bends the features back \cite{Zabolotnyy2013}.

The results of the tight-binding  fit are presented in Fig.3 (c) and Fig.3 (d). They clearly confirm the transition from hole-like anti-bonding (red contours) Fermi surface in OD 77K sample to electron-like one in OD 47K sample.  \\

In conclusion, a combined study of ERS and ARPES demonstrates that the PG collapses in between p=0.201 and p=0.236 as the FS change of topology from hole like to electron like in Bi-2212. This is in perfect agreement with a scenario for which the PG ends at a Lifshitz transition which occurs close to $p_c=0.22$ in Bi-2212 compound and that the PG only appears on a hole-like Fermi surface.

We are grateful to M. Civelli, A. Georges, I. Paul, and Louis Taillefer for fruitful discussions. Correspondences and requests for materials should be addressed to A.S. (alain.sacuto@univ-paris-diderot.fr); B.L. is supported by the DIM OxyMORE, Ile de France. S.V.B. is supported by DFG grant BO 1912/6-1. We acknowledge Diamond Light Source for time on beamline I05 under proposal SI17045.

\bibliographystyle{apsrev4-1}
\bibliography{referencesmain}

\end{document}